\documentstyle{mn}
\input{epsf}
\input rotate

\def\la{\lower.5ex\hbox{$\; \buildrel < \over \sim \;$}}
\def\ga{\lower.5ex\hbox{$\; \buildrel > \over \sim \;$}}
\def\lyal {{\rm Ly} \alpha}
\def\nh {N_{\scriptscriptstyle\rm HI}}
\def\j21{J_{-21}}
\def\tj21{\tilde{J}_{-21}}
\def\etal{et~al.\ }
  
\title
{On the enrichment of the intergalactic medium by galactic winds}
\author[Biman B. Nath and Neil Trentham]
       { Biman B. Nath$^1$\thanks{Present address: Raman Research
Institute, Bangalore 560080 India} and Neil Trentham$^2$\\
        $^1$Inter-University Centre for Astronomy and Astrophysics,
Post Box 4, Ganeshkhind, Pune -- 411007, India\\
$^2$Institute for Astronomy, 2680 Woodlawn Drive, University of Hawaii, Honolulu,
HI 96822, USA\\
(biman@rri.ernet.in, nat@ifa.hawaii.edu) }
\date{Accepted for publication in the MNRAS
}
 
\begin{document}
 
\maketitle
 
 \begin{abstract}
Observations of metal lines in $\lyal$ absorption systems of small
H~I column density and their ubiquitous nature
suggest that the intergalactic medium (IGM)
was enriched to about $Z \sim 0.01 \> $Z$_{\odot}$ by 
a redshift $z \sim 3$.
We investigate the role of winds from small star-forming galaxies 
at high $z$ in enriching the IGM. The existence of 
large numbers of small galaxies
at high $z$ follows naturally from hierarchical clustering theories (e.g.~CDM).

For analytical simplicity we assume that the
galactic winds escape the galaxies at a single
characteristic redshift $z_{in}$, and 
we model the galactic winds as spherical shock waves propagating
through the IGM. We then calculate the probability distribution of
the metallicity of the IGM, as a function of time (for different
values of $z_{in}$),
adopting plausible galaxy mass functions (from Press-Schechter
formalism), cooling physics, star-formation efficiencies, gas
ejection dynamics, and nucleosynthesis 
yields. We compare this expected
distribution with the observed distribution of metallicities in the
Ly$\alpha$ forest at $z=3$, the 
metal poor stars in the halo of our Galaxy,
and with other observational constraints on
such a scenario. We find that galactic winds
at high $z$ could have enriched the IGM to a mean metallicity of 
$Z \sim 0.01 $Z$_{\odot}$ 
at $z \sim 3$, 
with a standard deviation of the same
order, if $z_{in} \la 5$, and that this satisfies all the
observational constraints.

\end{abstract}

\begin{keywords}cosmology : 
early Universe --
galaxies : formation --
galaxies : abundances -- galaxies : evolution --
galaxies : intergalactic medium --
galaxies : quasars : absorption lines --
stars: Galactic halo --
stars: abundances -- 
stars: supernovae
\end{keywords} 

\section{Introduction}
Recent observations with the
Keck 10 m telescope have shown that relatively low H~I column 
density $\lyal$ absorption lines ($\nh \ga 10^{14}$ cm$^{-2}$) at $z \sim 3$
have corresponding metal lines. Cowie \etal (1995), Tytler \etal (1995)
and Womble \etal (1996) reported that more than half of all
$\lyal$ absorption systems with $\nh \ga 10^{14}$ cm$^{-2}$
have CIV. These observations confirmed previous suggestions of 
enrichment in the $\lyal$ forest clouds (Meyer \& York 1987; Tytler 1987).

More recently, Songaila \& Cowie (1996) have detected lines due to
several other ions such as SIV. They have also claimed that the
fraction of $\lyal$ lines with associated metal lines is larger than
$\sim 0.5$ as reported earlier. They found C~IV lines in
$90 \%$ of clouds 
with $N_{H~I} > 1.6 \times 10^{15}$ cm$^{-2}$ and $75 \%$ of clouds 
with $N_{H~I} > 3.0 \times 10^{14}$ cm$^{-2}$. They emphasize that
it would be difficult to estimate the exact fraction of $\lyal$
lines with associated metal lines with $\nh \la 10^{15}$ cm$^{-2}$
as it would mean detecting C~IV lines with $N_{C~IV} \la 10^{11}$
cm$^{-2}$ which is difficult at present. 
It is also of interest that Songaila \& Cowie (1996) found evidence
of enhancement of the alpha process elements versus the Fe process
elements, as is also the case in damped $\lyal$ systems (Lu \etal 1996), 
and in the metal poor stars of our Galactic halo. 

Such a large fraction of $\lyal$ lines with associated metal lines
suggests that the intergalactic medium at $z \sim 3$ was
enriched. The estimated  metallicity in the
$\lyal$ systems depends on the value of the ionization parameter assumed.
Cowie \etal (1995), Womble \etal (1995) and Songaila \& Cowie (1996)
estimated the metallicity to be of order $Z \sim 10^{-2} \> $Z$_{\odot}$
for an ionization parameter in the range $\Gamma \sim 10^{-1.5}\hbox{--} 
10^{-2.5}$. Rauch, Haehnelt \& Steinmetz (1996), however,
 compared the result of their numerical simulation with the observed
data and estimated a mean metallicity of $Z \sim 10^{-2.5} \> $Z$_{\odot}$.

The IGM could have been enriched in a number of ways,
e.g., Population III stars, early-forming
very massive (10$^{2}$ to 10$^{5}$ M$_{\odot}$) 
objects (Carr, Bond \& Arnett 1984)
or galactic winds from galaxies at high redshifts. In this
paper, we investigate the possibility that supernova-driven galactic winds
from small star-forming galaxies
at high redshift could have enriched the IGM. 
Such low-mass galaxies form naturally at high $z$ in hierarchical
clustering scenarios for galaxy formation, like CDM.  If the
gas in these galaxies can cool and form stars, then any metals
produced when these stars go supernova can easily be ejected from
the galaxy by supernova-driven winds, as the galaxies have shallow 
gravitational potential wells (Dekel \& Silk 1986).  
This was in essence the scenario 
outlined by Couchman \& Rees (1986), who consider the subsequent
influence of these galaxies on structure formation.  Here we
investigate this scenario in detail, incorporating up-to-date
cooling physics and compare our models to recent observations,
concentrating on issues most directly related to metal production.

The amount of gas
and the metallicity of the ejected gas from each
galaxy depends on the mass of the galaxy
and the parameters of star formation (e.g., Nath \& Chiba 1995). The
enriched gas would be carried in cooled shells which will propagate through
the IGM, sweeping up matter. The shells (each with its own radius
and metallicity, which have been fixed by the mass of the parent
galaxy) 
will then collide and the enriched gas will
be mixed. Given a long time, the gas will distribute 
more homogeneously and
the resulting metallicity will depend on the mass function of galaxies,
the redshifts when the galaxies erupted and the parameters of star formation
(in addition to the cosmological parameters).

Several authors have commented on the production of metals 
as a possible by-product of astrophysical processes
occurring during the early stage of galaxy formation
(e.g.~Fukugita \& Kawasaki 1994, 
Voit 1996).
Specifically, if supernovae are partly responsible for the
generation of the UV background (particularly at soft
energies; see Miralda-Escud\'e \& Ostriker
1990, Madau \& Shull 1996) and/or the reionization of the Universe
(see Giroux \& Shapiro 1996),
then simultaneous cosmic metal production seems inevitable.
Recently, Gnedin and Ostriker (1997) have presented the results
of their numerical simulations of reionization and metal production
at high redshift by Population III stars. Miralda-Escud\'e and Rees
(1997) have discussed the production of metals by high redshift
supernovae, but did not consider the details of the gas ejection
dynamics and the mixing of the gas in the IGM.
Even prior to these works, Silk, Wyse \& Shields (1987) had
contemplated on the enrichment of the IGM by bursting dwarf galaxies
as a plausible cosmological process, continuing to the present day, 
along with the generation of the soft UV background. 

This paper is structured as follows.
In Section 2 
we outline the chemical history of the Universe from the Big Bang to the
present day, and place our model in this context and in the more
general context of galaxy formation theories.
In Section 3 
we estimate the time evolution of the mean metallicity
of the IGM.  We also estimate its variance, from the details
of the collisions between the shells and the consequent mixing
of the gas. For simplicity, we assume a single epoch at 
which the galactic
winds go off, assume  the shocks to be spherical, and then model the 
evolution of the probability distribution of the IGM metallicity.
We then compare this distribution with the observed metallicities of
the Ly$\alpha$ clouds at $z=3$, the metallicity distribution of the
metal-poor stars in the halo of our Galaxy, and other observations in
Section 4.

\bigskip
\section{The chemical history of the Universe}

In this section we outline a plausible 
scenario that describes the chemical history
of the Universe.  We do this so as to place the model we present
here in the wider framework of the galaxy formation problem
and also to highlight the
observational contexts in which our results are and are not valid.

After the Big Bang, only hydrogen, helium, and trace amounts of lithium were 
present in the Universe; carbon and larger nuclei essentially did not exist.
As the initial fluctuations in the dark matter spectrum 
grew, at some redshift, the very high 
$\sigma$ peaks reached masses large enough
that the baryonic component contained within these
peaks may cool through molecular transitions  
(Blumenthal \etal 1984, Couchman \& Rees 1986, Tegmark et al.~1997).
The relevant baryonic 
mass scales vary from 10$^3$ to 10$^7$ M$_{\odot}$ for
$100 > z > 10$ (see Figure 6 of Tegmark et al.~1997).
The molecular hydrogen responsible for triggering
the cooling mostly comes from H$^{+}$+H$^{-}$ interactions,
where the H$^{-}$ is made from H+e$^{-}$ reactions (a small
number of free electrons remain after recombination; Couchman
\& Rees 1986).  
No metals exist yet to assist in the cooling (however, see
Gnedin \& Ostriker 1992 for a possible alternative).
After collapse is initiated and the densities get 
high H+H+H reactions presumably take over as the primary source
of making H$_{2}$.  No dust grains are available so that absorption
of H~I
onto dust, which is the dominant method of making H$_2$ in 
star-forming regions in the
Milky Way, is not important.  If sufficient H$_2$ is made, the
cooled baryons may form stars.  If this happens, then some of these
stars might become
supernovae and produce some metals and UV photons.  But at very high
$z >> 10$, these high $\sigma$
objects are extremely rare, so that they do not produce
significant amounts of metals (they may, however, reheat
the IGM; Tegmark et al.~1997).
 
As the age of the Universe increases, the much more common lower $\sigma$
peaks reach the mass scales required for the gas to cool, and baryonic
collapse and possibly star formation becomes much more ubiquitous.
It is this phase that we are concerned with in this paper.
The mass scale here will be set mostly by atomic cooling,
as molecular hydrogen will easily be photodissociated by the
few UV photons produced by the few higher $\sigma$ peaks (Haiman, Rees,
\& Loeb 1996).
The main uncertainty 
preventing a rigorous study of this
is the stellar initial mass function (IMF) in these galaxies,
where the collapsing gas is essentially unenriched.  We know that it
is probably more biassed to high masses than the local Salpeter IMF
because otherwise at least 20\% of halo stars would be pregalactic and 
made out of material that has only been processed once (i.e.~during
this phase; Miralda-Escud\'e \& Rees, 1996).  
If this were true, then at 
least 20\% of the halo stars would be heavy neutron-capture deficient
(see Truran 1981, also Section 4.2), which is unlikely 
(Ryan, Norris \& Beers 1996).  In this work, we adopt a parameterization
that takes this lack of knowledge 
about the IMF into account. 

Here we adopt plausible
galaxy mass functions (from Press-Schechter (1974) theory), cooling
efficiencies, and nucleosynthesis yields to compute the amount of
metals produced.  We then model the propagation of these metals
by galactic winds through the IGM and compute the mean metallicity 
of the IGM and its spatial variance as a function of time.  
The mean value we get, about 0.01 Z$_{\odot}$, compares well with 
the metallicities of the Ly$\alpha$ clouds seen at $z=3$, and we
conclude that the metals seen there 
could result from this phase of cosmic
enrichment.  
In such a scenario,
local star formation within the Ly$\alpha$ forest
clouds themselves does not contribute many of the metals.
This would not be surprising, as the systems
are very small and 
diffuse (on the other hand, in the bigger, denser, and more
metal-rich damped Ly$\alpha$ systems, self-enrichment almost certainly
happens).

It is also probable that many of the lowest $Z$
halo stars, in particular the Sr-deficient stars, might have
been  made from material that was only processed during this phase
(see Section 4.2).
Most of the rest of the halo probably came from material that was
enriched by subsequent generations of stars that formed either in
smaller stellar systems that were the Milky Way progenitors, or
during the collapse of the halo (Eggen, Lynden-Bell \& Sandage
1962).
Another important consequence of the
enrichment process is the generation of a large
number of UV photons, which will raise the Jeans mass of the IGM,
hence inhibiting further collapse of baryons into dark matter halos
(e.g., Thoul \& Weinberg 1995).  
What this means is that the redshift range over which this phase
happens may be quite small (in this work we
approximate it by a single characteristic value $z_{in}$).
These UV photons may
also in whole or part reionize the IGM (depending on the
efficiency with which they can escape the galaxies), therefore
explaining the H~I Gunn-Peterson effect.  

After this phase,  
star formation continues to happen in galaxies, although it may
well have been supressed (particularly in small galaxies;
see Babul \& Rees 1992, Efstathiou 1992)
at $z>2$ by the UV background produced by
the galaxies described above, and by quasars at high redshift.
Large numbers of star-forming galaxies are observed  in the field
(see e.g.~Broadhurst, Ellis \& Shanks 1988; Cowie, Songaila \& Hu 1991), 
but these are typically more massive than the galaxies that are
relevant in the preceding two paragraphs (Cowie, Hu \& Songaila 1995).
Therefore any metals produced are unlikely to escape 
the galaxies because of
the deep potential wells.   
Galaxies like the Milky Way have also experienced considerable
star formation in the disk and have retained the metals produced;
this is why the disk stars and the local interstellar
medium (ISM) have metallicities
close to solar, much higher than the 0.01 Z$_{\odot}$ produced
in the phase above.
Some of these galaxies may have since
been tidally disrupted in 
interactions and the metals transferred from
their ISM into the IGM.  Because these galaxies might
have had fairly metal-rich ISMs the IGM metallicity at $z= 0$
might be somewhat higher than 0.01 Z$_{\odot}$; strong observational
constraints on the $z=0$ IGM metallicity do not exist
at present. However, the Milky Way formed out of a gas cloud
that was probably close to 0.01 Z$_{\odot}$ in mean metallicity
(this is inferred from the peak of the halo metallicity distribution
function of Ryan \& Norris 1991), so that the IGM was not 
probably significantly enriched above this value by the time the
Milky Way formed.

The one case where the IGM metallicity at $z=0$ is well known,
is of course in galaxy clusters, where the hot intracluster
medium (ICM) is metal-rich, with $Z$ of 0.1 to 1 Z$_{\odot}$.
The measurement of the iron mass to optical luminosity ratio
suggests that the cluster ellipticals provided these metals
(Arnaud et al.~1992).  The detection of Type II abundances with 
$ASCA$ suggests that the metals come from Type II supernovae,
which are not observed to occur in present-day ellipticals
(Mushotzky et al.~1996, Loewenstein \& Mushotzky 1996).  This
suggests that these
ellipticals might have undergone a starburst phase early
in their history (possibly an ultraluminous phase; see
Sanders \& Mirabel 1996), where substantial amounts of metals were made
and which have since entered the ICM by cluster-related processes
like ram-pressure stripping.  

\bigskip
\section{Galactic winds and the IGM}

We will begin by explaining the different building blocks of our model ---
the galactic winds, their propagation in the IGM, and the interactions
between the shells --- and then compute the time evolution of the metallicity
distribution.

\subsection{Galactic winds}

Galactic winds are believed to originate when the thermal
energy of the interstellar gas in a galaxy
exceeds its binding energy. The amount of
gas and metals ejected by a galaxy of mass (including non-luminous matter)
$M$ depends on the parameters of star formation, {\it viz.}, the 
IMF, star formation rate, and the efficiency of supernovae
remnants in heating the ISM gas (Yoshii \& Arimoto 1987). A variety of
numerical works calculated the time for initiating the galactic wind
and the metal enrichment as a function of these parameters (see, e.g., 
Matteucci \& Gibson 1995 and references therein). We will use the
results of Nath \& Chiba (1995), which are based on simple assumptions,
but are close to those obtained from more sophisticated approaches,
for our purpose.

Nath \& Chiba (1995; hereafter NC95) estimated the fraction $\gamma$
of the galactic mass $M$ ejected in a wind ($M_{wind}=\gamma M$) (their
equations (10) and (14)), 
and the metallicity of the ejected gas, $Z_{wind}$ (their eqn (9)), using
iron as the trace element. The total amount of metals
ejected from each galaxy is then 
$Z_{wind} M_{wind}$. 
These values depend on
the slope of the IMF ($\phi(m) \propto m^{-x} dm$, $x=1.35$ for Salpeter
IMF) and
the efficiency of the supernova remnants in heating up the ISM
gas ($\eta \sim 0.1$). They incorporated the IMF slope, the upper and
lower limits of masses of main sequence stars and the lower limit of
the mass of stars which go supernova, into a single parameter $\nu_{50}$,
which was defined as the number of supernovae per $50$ M$_{\odot}$ of
baryons. For a Salpeter IMF, $x=1.35$, $\nu_{50} \sim 0.37$ (for details
see NC95).
The important concept here is that $\gamma$ and $Z_{wind}$ are
single-valued functions of $M$ given $\nu_{50}$ and $\eta$
(typical examples of these functions are given in Figure 1 of NC95).
Consequently, the functions $\gamma (M)$ and $Z_{wind} (M)$, combined
with the mass function of the galaxies, yield the total
amount of metals produced. 

The total energy in the explosion (meaning the
cumulative effect of all the supernovae in a galaxy)
is uncertain and we adopt the method of
Tegmark, Silk, \& 
Evrard (1993, hereafter TSE93) to estimate this.  
We need to know this number because it determines how
well the winds and the metals they carry are propagated
through the IGM.
Since the galactic winds
last for $t_{burn} \sim 5 \times 10^7$ yr, which is much smaller than any other
timescales in this calculation, we assume them to be ejected in an explosive
manner. Making this assumption, the total energy in the galactic wind is then
$0.02 \times 0.007 \times M_b c^2$, where $M_b$ is the baryonic mass in the
galaxy. The 0.02 comes from the observations of Heckman, Armus,
\& Miley (1990) who show
that 2\% of the total luminosity of high-z starburst galaxies goes into
galactic winds. If one assumes the galactic wind to have a constant mechanical
luminosity $L_{sn}$ for the duration of $t_{burn}$, then TSE93 show that 
$L_{sn} \sim (M_b/M_\odot) L_\odot$. In the calculations below, we
assume $M_b=0.1 \> M$ (we discuss the effect of the uncertainty in this
relation on the final results in \S 4).

\subsection{Propagation of galactic winds in the IGM}

 We will again follow TSE93  
in calculating the propagation of the galactic winds
in the IGM. The matter in the wind is assumed
to be in a dense, cool
shell (except for a small fraction of mass in the hot interior), 
that is 
sweeping more mass from the IGM. The 
density of particles in the shell can be estimated by equating the
ram pressure ($m_p n_a V_{sh}^2$, where $n_a$ is the ambient 
particle density and $V_{sh}$ is the shock velocity) to the thermal
pressure of the gas in the shell, which is assumed to be cold (at 
$T \sim 10^4$ K, the lowest temperature attained by gas in the absence
of molecular cooling). This gives the particle density in the shell,
$n_s = n_a \> 120 \> V_{sh, 100}^2$, where the shock velocity is written
in the units of $100$ km/s. Assuming the total mass inside the radius
$R$ of the shell to be contained 
within the shell, the thickness of the shell is
$\delta R = (R/360) V_{sh,100}^{-2}$.

Let the pressure inside the shell be $p$, which is driving the shell
outward with a force $4 \pi R^2 p$. The sweeping of matter from the IGM
provides a braking force equal to $(\dot R - HR) {dm \over dt}=
(\dot R -HR) \times 3 ({{\dot R} \over {R}} -H)$, 
where $H$ is the Hubble constant.
There is another braking force due to the gravitation of mass in the shell.
If the dark matter density (in units of the critical density) is $\Omega_d$,
then the equation of motion of the shell is (eqn (1) of TSE93) given by
\begin{equation}
\ddot R= {8 \pi p G \over \Omega_{IGM} H^2 R} - {3 \over R} (\dot R -HR)^2
-  ( \Omega_d + 0.5 \Omega_{IGM}) (0.5 H^2 R) \>.
\end{equation} 
Here, $\Omega_{IGM}$ is the ratio of the density of the intergalactic
gas to the critical density. This should be differentiated from the
total baryon density in the universe, $\Omega_b$ (expressed in the units
of the critical density), as some of the baryons would be in the collapsed
halos and not in the intergalactic medium.

The evolution of the total energy inside 
the shell ($E_t={3 \over 2} p V=
2 \pi p R^3$) is given by (TSE93)
\begin{equation}
{d E_t \over dt}= L_{sn}-4 \pi p R^2 \dot R - L_{brem} -L_{comp}.
\end{equation}
Here, the first term refers to the energy input due to supernovae,
the second term is due to adiabatic loss, the third term describes
energy loss due to free-free radiation and the last term is
the energy loss rate due to inverse Compton scattering off the microwave
background photons. 

We solve the above two equations for the propagation of the shells
in the IGM. For the initial conditions, we assume that the initial
radius of the shell is the radius of the galaxy.
We adopt the scaling of Saito (1979) for the typical size of a
galaxy ($R=1.2 \times 10^2 (M/ 10^{12} M_{\odot})^{0.55}$
kpc). The initial velocity is expected to be of the order of the thermal
velocity of the ISM gas which is presumably heated to $\sim 10^6$ K
(Babul \& Rees 1992),
i.e. $V_{sh, 100} \sim 1$ initially. However, we found that the
results are almost independent of the initial velocity, as the kick
due to $L_{sn}$, which we assume to continue for $5 \times 10^7$ yr, 
basically determines the propagation of the shell. 

\begin{figure}
\protect\centerline{
\epsfysize=3.5in\epsffile[30 160 570 700]
{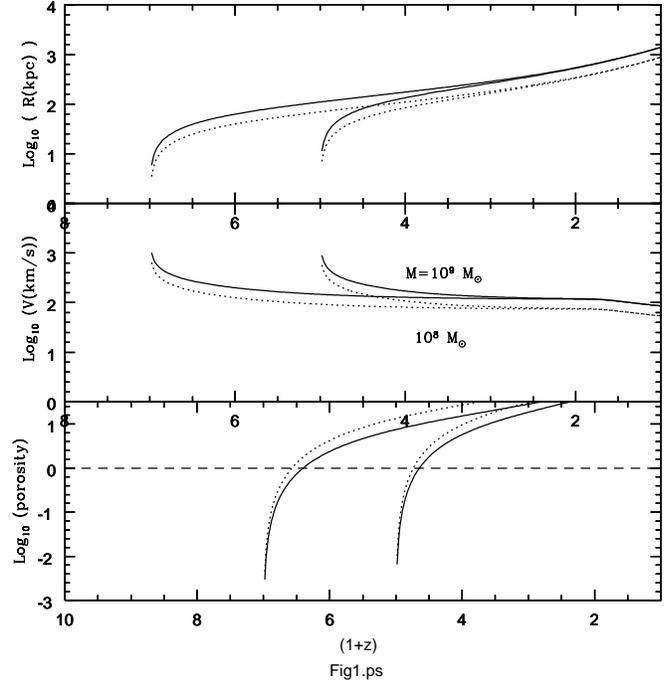}
}
\caption{ The uppermost panel shows the evolution
of the radii of the shells from parent galaxies of mass
$M=10^8$ M$_{\odot}$ (dotted) and $M=10^9$ M$_{\odot}$ (solid),
for $1+z_{in}=5$ and $7$.
The middle panel shows the corresponding shell velocities. The
lowest panels shows the evolution of the porosity. The calculations
assume sCDM model and $\Omega_{IGM}=0.05$.}
\end{figure}

The upper two
panels of Fig.~1 show the evolution of the radius and velocity of the 
shell ejected from
galaxies with masses $10^8$ and $10^9$ M$_{\odot}$ at 
redshifts $1+z_{in}= 5$ and 7. 

\subsection{Porosity}

If the comoving mass function of the parent objects
which ejected
their galactic winds at a certain redshift, $z_{in}$, is $n(M,z_{in})$
(which has the dimensions of Mpc$^{-3}$ $M^{-1}$)
then the porosity $por$ of the IGM at some later redshift can be defined as,
\begin{equation}
{d \over d M} 
por(M, z, z_{in})={4 \pi \over 3} R(M, z, z_{in})^3 \> 
n(M,z_{in}) \>
\Bigl ({1+z \over 1+z_{in}} \Bigr )^3 \>.
\end{equation}
The porosity here simply denotes the filling factor of the spherical shocks
(volume $\times$ number density of parent objects) for $por
\le 1$. For larger values of $por$, the shells overlap one another,
and in this case, the filling factor of the {\it overlapped regions}
is $1-por^{-1}$.
The function $n(M,z_{in})$ can be estimated for a given model of structure
formation from Press-Schechter theory. 

The Press-Schechter mass function only refers to objects with a density
contrast above a threshold value.  To be specific, it counts objects
with linearly extrapolated overdensity $\delta (1+z_c) > 
\delta_c (1+z_c)$, where $z_c$ is the collapse redshift. To form a galaxy 
by a given $z$ the gas in the halo must, in addition to
reaching some threshold density contrast, cool within a Hubble time
$t_{H} (z)$.  We take this effect into account using
the approach of Peacock
\& Heavens (1990). They showed that the effect of cooling can be taken
into account by changing $(1+z_c)$ in the Press-Schechter mass function
to $(1+z)(1+M/M_{cool})^{2/3}$, for galaxies to have formed by a given
redshift $z$, where
\begin{equation}
M_{cool} \sim 3.6 \times 10^{11} \> ({\Omega_b \over 0.05}
) \> M_{\odot} \>;
\end{equation}
here we use 
the nucleosynthesis value of $\Omega_b \sim 0.05 \> h_{50}^{-2}$
($h_{50}$ is the Hubble constant in units of 50 km s$^{-1}$ Mpc$^{-1}$)
and we have assumed cooling in the absence of metals. We use this
function for our calculations below.

Formally, we should extend the above calculations to include the constraint
that not only should the galaxy form, but it should also have 
had enough time
for the supernovae to heat up the ISM and for the galactic wind to ensue.
However, we find that, for reasonable parameters of star formation
($\nu_{50}=0.35, \eta=0.1$), the ratio of the time scale for 
galactic winds ($t_{gw}$, see NC95)
and the cooling time of the gas $t_{cool}$ is
\begin{equation}
{t_{gw} \over t_{cool}} \sim 2 \> \Bigl ( {M \over 10^9 \> M_{\odot}}
\Bigr )^{-0.225} \Bigl ( {1+z_c \over 6} \Bigr )^{3/2} \>,
\end{equation}
where we have assumed primordial cooling (Peacock and Heavens 1990),
$\Omega_b=0.05$ and $h_{50}=1$. The two time scales are
of the same order for the range of masses we are interested in. Therefore,
the additional constraint $(t_{cool}+t_{gw}) \leq t_H$ will not substantially
change the above mass function, and so we neglect it here. 

As a template model of structure formation, we use the standard CDM (hereafter
sCDM) model with $h_{50}=1$ 
and $\Omega=1$. We use the normalization of $\sigma _8
=1.2$ from the analysis of COBE data of four years (Bunn \& White 1996). 
We use the analytical fit to 
the CDM power spectrum as given in White \& Frenk (1991) for 
our calculations.

In the lower
panel of Fig.~1,
we plot the porosity as a function of redshift for $M=10^8$ and 
$10^9$ M$_{\odot}$ for $1+z_{in}=5,7$ in the sCDM model, calculated by
integrating eqn ($3$) with $n(M^{\prime}) \propto \delta(M^{\prime} - M)$. 
The important thing to note in the figure is that shells from more
massive galaxies reach the unit porosity limit later in time.

\subsection{Metallicity of the shells}

The metallicity of the shells will depend on (a) the amount of metals
ejected by the parent galaxy ($M_{wind} Z_{wind}$; see above in \S 3.1), (b)
the total amount of gas ejected ($M_{wind}$), and (c) the amount of
IGM gas swept up by the shell ($(4 \pi /3) R^3 \rho_c \Omega_{IGM}$,
assuming that it has not collided with
any other shells so far; otherwise see below). 
The metallicity of the shell in this case is then given by
\begin{equation}
Z_{shell}={M_{wind} Z_{wind} \over M_{wind} + 
(4 \pi /3) R^3 \rho_c \Omega_{IGM} }
\>.
\end{equation}

If the enrichment of the IGM is absolutely homogeneous, then the total
metallicity of the IGM will be
\begin{equation}
Z_{IGM, hom} (z_{in})= {\int M_{wind} Z_{wind} n(M, z_{in}) dM \over
\int M_{wind} n(M, z_{in}) dM + \rho_c \Omega_{IGM} } \>.
\end{equation}
Here the integration is over the range of galactic masses.
We have used a lower limit of $10^6$ M$_{\odot}$ in the calculations.
The value of $Z_{IGM,hom}$
  will depend on the epoch of galactic winds $z_{in}$, the mass
function of galaxies and the value of $\Omega_{IGM}$, in addition to the
parameters of star formation in the parent galaxies. For sCDM models,
and for $\Omega_{IGM}=0.05$, we find that $Z_{IGM,hom}/ Z_{\odot}
=0.033, 0.045, 
0.057$, for $1+z_{in}=10, 7, 5$ respectively.
(These numbers do not change much for other values of $\Omega_{IGM}$;
for $\Omega_{IGM}=0.01$, for example, they are practically the same.)
However, the state of the IGM is unlikely to be that with a constant
metallicity, due to the inefficient mixing process. We next discuss the
mixing due to collisions between shells. But before we do so, let us
define another quantity, the homogeneous metallicity of the IGM due
to galactic winds from galaxies of mass lower than a certain value $M'$,
\begin{equation}
Z_{IGM, hom} (M' <M, z_{in})= {\int ^{M} M_{wind} Z_{wind} n(M', z_{in}) 
dM'  \over
\int ^{M} M_{wind} n(M', z_{in}) dM' +  \rho_c \Omega_{IGM} } \>.
\end{equation}

\subsection{Interaction between shells}

When the shells meet, the result of 
each interaction will depend on the 
velocity of the shells
involved. With the assumption of a single $z_{in}$ for all
galactic winds, one can make a one-to-one correspondence between the velocity
of the shell and the mass of the parent galaxy, and this makes it simpler
to trace the evolution of the shells, as explained below.

The interactions between the shells are very complicated in general. It will
be useful here to divide them into two categories. The first type of
interaction is that between shells of different sizes. For simplicity, let
us assume that the interacting shells differ greatly in their sizes
($R_1 \ll R_2$, at some redshift $z$). 
In the approximation of a single epoch of galactic winds,
as the curves in Fig.~1 show, 
the corresponding velocities are then
also very different, with $V_1 \ll V_2$.
It is then also true (from the same curves) that the masses of the parent
galaxies and therefore the  metallicity of the shells are different:
$Z_{shell, 1} \ll Z_{shell, 2}$.

We assume that in this case, the shell with larger velocity and
momentum (and therefore,
with higher metallicity) will sweep the matter in the other shell with it.
Let us look at the shells of size $R_2$. Shells of much smaller size 
$R_1 \ll R_2$ would
have attained unit porosity much earlier than these shells. In other words,
by the time the shells of
size $R_2$ have reached a reasonably large size, shells
of much smaller sizes would have overlapped and distributed their metal
enriched gas 
homogeneously in the IGM. For shells which have reached a size $R_2$ 
at some redshift, the IGM gas already has a metallicity corresponding
to shells of smaller sizes, originating from smaller galactic masses,
that is, $Z_{IGM, hom} (M'<M_2, z_{in})$, where 
$R_2$ is the size of the shell from
a parent galaxy of mass $M_2$. 
Although, strictly speaking, this is true only for $R_2 \gg R_1$, for
analytical simplicity we will assume it to hold also for $R_2 > R_1$.
We, therefore, modify our previous expression for the metallicity of the
shells from galaxies with mass $M$ (eqn $6$),
and write
\begin{eqnarray}
&&Z_{shell, M}=\nonumber\\
&&{M_{wind} Z_{wind} + (4 \pi /3) R^3 \Omega_{IGM} \> \rho_c 
\> Z_{IGM, hom} (M' < M, z_{in}) \over M_{wind} + 
(4 \pi /3) R^3 \rho_c \Omega_{IGM} }
\end{eqnarray}
(For future reference, let us call this assumption {\bf A1}.)
This function $Z_{shell, M}$ is a monotonically
increasing function of $M$, a fact which will be useful to us below. 

We have seen (in \S 3.3) that shells from
more massive galaxies reach the unit porosity limit later in time. Consider
now the shells corresponding to mass (of parent galaxies) $M_{\ast}$. Once
the porosity of these shells reach unity, the IGM gas, including the
gas and metals from shells corresponding to $M < M_{\ast}$, has been 
more or less raked up by these shells. This means that 
$Z_{shell, M_{\ast}}$ does not increase any longer. Therefore,
we freeze the metallicity of the shells once they reach unit porosity,
and keep it constant during their evolution later.

\begin{figure}
\protect\centerline{
\epsfysize=2.5in\epsffile[25 160 570 700]
{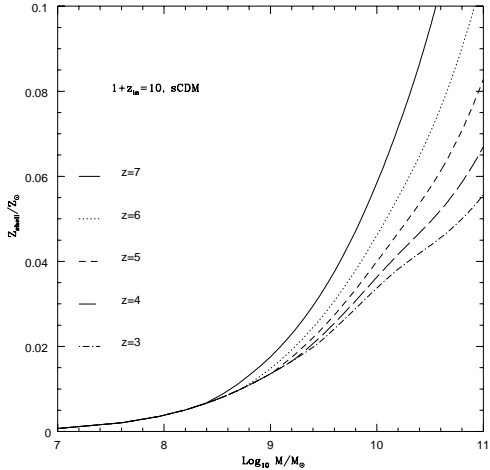}
}
\caption{ The shell metallicity, $Z_{shell, M}$, is
shown as a function of the galactic mass $M$ at different redshifts,
for $1+z_{in}=10$, sCDM model and $\Omega_{IGM}=0.05$.}
\end{figure}

Fig. 2 shows the curves for $Z_{shell, M}$ as a function of $M$ at 
different redshifts, for $1+z_{in}=10$. In general, the shell metallicity
decreases in time because of sweeping more and more of IGM gas,
which must have lower $Z$. At later
times (small $z$), the curves, therefore, flatten out. Note that,
after reaching unit porosity, the shell metallicity does not change, 
giving rise to the envelope of the curves at the left hand side of 
Fig. 2. 

The second type of interaction takes place between shells of similar
sizes ($R_1 \sim R_2$). This type of collisions in general
gives rise to two
reflected shockwaves moving away from the plane of intersection, two contact
discontinuities and another shock wave forming an annulus where the
shells intersect. For simplicity, we neglect the annulus here.
Furthermore, for geometrical simplicity, we assume that the
region bounded by the reflected shockwaves can be reproduced by
imagining the colliding shells passing through each other, since
the reflected shocks have similar velocities as the original shells (see, e.g.,
Voinovich \& Chernin 1995).
The matter in this region
is compressed and heated twice by the reflected shocks to a temperature
$\sim 6 \times 10^4$ K, for a typical shell velocity $V_sh \sim 50$
km/s (see Fig. 1). The sound velocity of the gas at this temperature
is $\sim 30$ km s$^{-1}$.
We assume that this gas fills up homogeneously the region bounded
by the reflected shocks (or, in our oversimplified picture, by the
shells continued {\it through} each other). This is shown schematically
in the right panel of Fig. 3. Let us call this assumption {\bf A2}.
We note here that, this is
strictly true only for shocks which do not have cold shell; in the case of
cold shells, the gas forms a ring (Yoshika
\& Ikeuchi 1990). 

\subsection{The probability distribution of metallicity}

Let us consider the galaxies of mass $M$. At some redshift $z < z_{in}$,
the galactic wind shells from these galaxies have radii $R$. If the 
porosity for only these shells (as defined in \S 3.3) is less than
unity (Case I in Figure 3), 
then these shells will have mostly collided with shells of
smaller size, which are more numerous than shells of size greater than
or equal to $R$. The shell will therefore have a metallicity $Z_{shell, M}$ 
given
by eqn (9) above (because of assumption {\bf A1}). Since the gas is accumulated
in the shell of thickness $\delta R$, the filling factor of the gas
of metallicity $Z_{shell, M}$ is $\lbrack (3 \delta R/ R) \> por(M, z, z_{in}) 
\rbrack$. We could
tentatively equate this as the probability of the IGM gas to have
a metallicity $Z=Z_{shell, M}$. However, because of assumption {\bf A1},
we note that the shells of size $R$ will also be pushed around by
shells of larger size, and therefore, be mixed with higher metallicity gas
(since $Z_{shell, M}$ is an increasing function of $M$, or equivalently $R$; 
see Fig. 2). 
Strictly speaking, then, the above filling factor will equal
the probability that the intergalactic gas has metallicity $Z$
or higher. Let us call this quantity $p(>Z)$.

\begin{figure}
\setbox1=\hbox{\epsfxsize=2.5in \epsffile{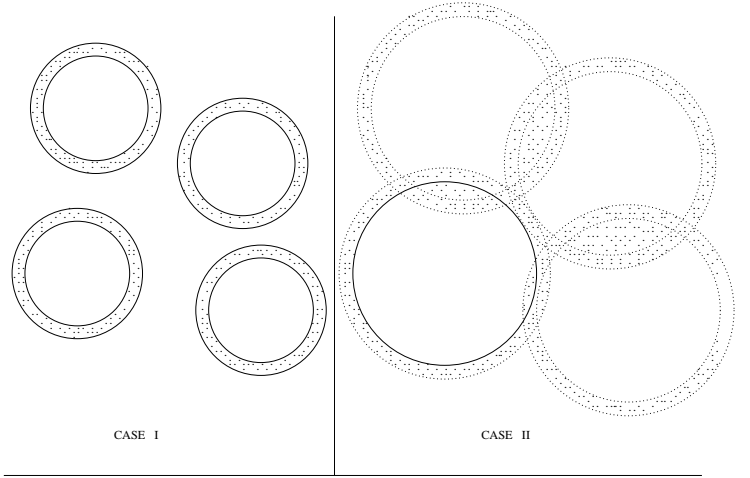}}
\centerline{\rotl1}

\caption{The two limits of calculating the filling
factor of gas of a given metallicity are shown schematically. In
Case I, the gas with the metallicity $Z_{shell, M}$ is confined within
the shells originating from galactic mass $M$. In Case II, the gas
occupies the overlapping regions.
Note that the structure
of shocks in the interacting region is very simplified here (see text).}
\end{figure}

If the porosity is larger than unity (Case II in Figure 3), then the filling
factor of the gas with metallicity $Z=Z_{shell, M}$ at redshift $z$ is simply
$(1- (1/por (M, z, z_{in})))$, from assumption {\bf A2}. Again for the above 
reasons,
we will equate this to $p(>Z)$.

We can then consider galactic winds from galaxies of different
masses, calculate the shell metallicities $Z=Z_{shell, M}$ at each
redshift, and calculate the function $p(>Z)$ using the algorithm described
above, depending on the value of the porosity for these shells. This function
is essentially $p(>Z)= \int_0^{Z} p^{\prime}(Z^{\prime}) dZ^{\prime}$,
where $p^{\prime}(Z^{\prime})$ is the probability distribution function of
the IGM metallicity.

\bigskip
\section{Results and discussions}

We plot in the left panels of
Fig 4a, 4b and 4c, the function $p(>Z)$ at various redshifts
for the cases $1+z_{in}=10, 7, 5$. We assumed $\Omega_{IGM}=0.05$,
$\nu_{50}=0.37$, $\eta=0.1$, and sCDM model (with
$\Omega=1$, $h_{50}=1$). The right panels show the the mean
and the standard deviation as a function of the redshift for the
three cases. The dashed line in the right panels show the values
of $Z_{IGM, hom}$, {\it i.e.}, the metallicity of the IGM if the
metals were distributed homogeneously. It is seen that the mean metallicity
increases in time to approach this value.

\begin{figure}
\protect\centerline{
\epsfysize=3.5in\epsffile[25 150 570 700]
{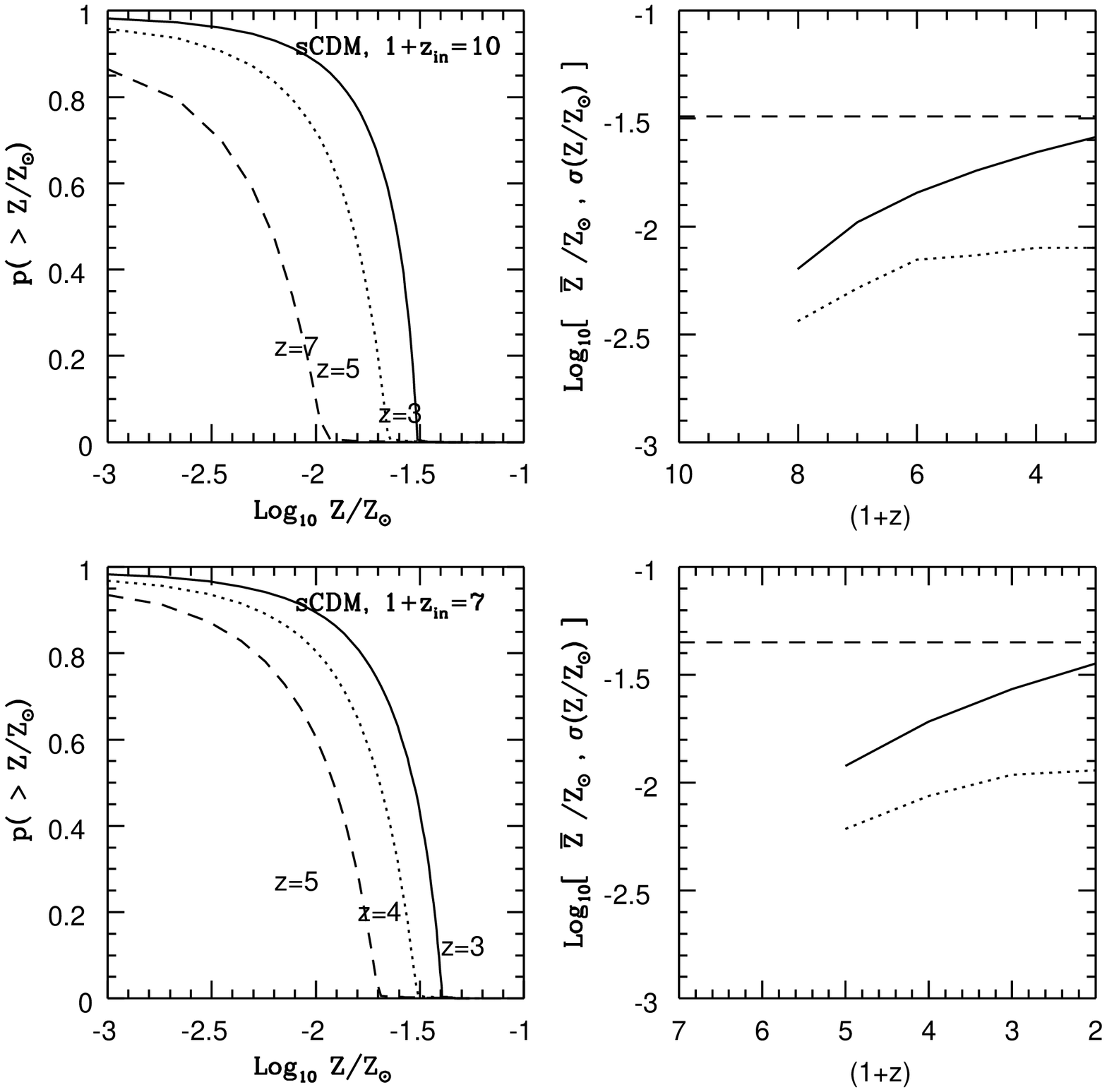}
}
\protect\centerline{
\epsfysize=2.2in\epsffile[25 350 570 700]
{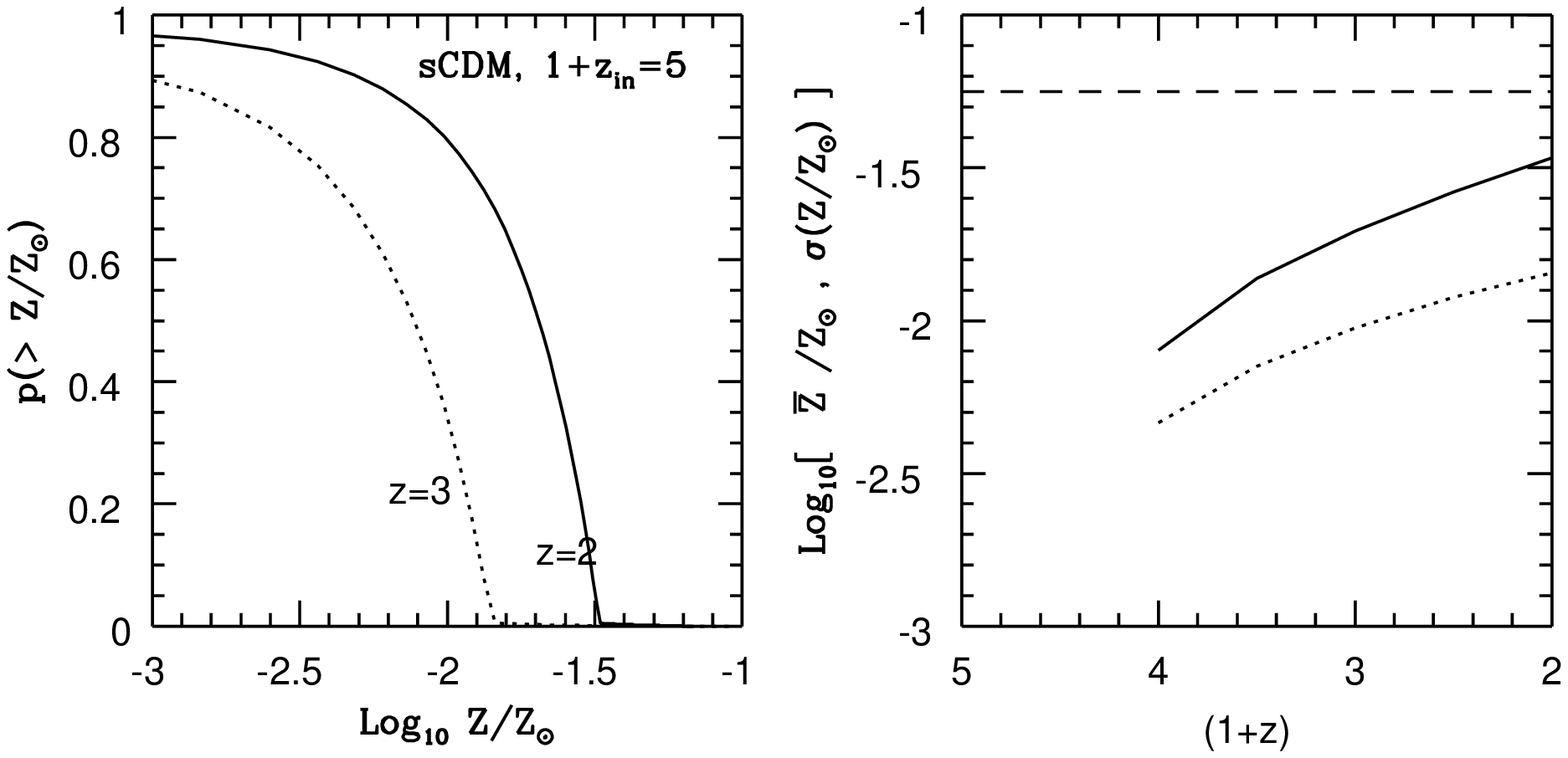}
}
\caption{{\bf (a)} In the left panel,
the function $p(>Z)$
is shown at redshifts $z=7,5,3$ for the case of $1+z_{in}=10$.
In the right panel, the mean (solid line) and the standard
deviation (dotted line) of the IGM metallicity
are shown as functions of redshift.
The dashed curve shows the metallicity had the gas from the
galaxies distributed in the IGM homogeneously. The curves
are for sCDM model and $\Omega_{IGM}=0.05$.
\noindent
{\bf (b)} The left panel shows $p(>Z)$ at $z=5, 4, 3$ for $1+z_{in}=7$.
The right panel shows the evolution of the mean (solid line)
and the standard deviation of the IGM metallicity.
\noindent
{\bf (c)} The left panel shows $p(>Z)$ at $z=3$ and $2$ for $1+z_{in}=5$.
The right panel shows the evolution of the mean (solid line)
and the standard deviation of the IGM metallicity.}
\end{figure}

A note on the uncertainties involved in the calculations
is in order here. We have done the calculations for other values of
the parameters $\Omega_{IGM}$ and $L_{SN}$. 
Changing the value of $\Omega_{IGM}$ to $0.01$,
increases the shell metallicities,
and shifts the curves horizontally by at the most $0.3$ dex. Decreasing
$L_{SN}$ by a factor of two, decreases the metallicities by $\sim 0.15$ dex.

\subsection{Comparison with the metallicity of $\lyal$ systems}

The reported metallicity of the $\lyal$ systems at $z \sim 3$ is
$Z/Z_{\odot} \sim 10^{-2.5} \hbox{--} 10^{-2}$ (Rauch \etal 1996,
Cowie \etal 1995). 
Fig 5 shows the mean metallicity and the standard deviation at $z=3$
in our model, as functions of $z_{in}$. Curves are shown for $\Omega_
{IGM}=0.01$ and $0.05$. The figure shows that
our model conforms to the observed
values of the mean metallicity of the $\lyal$ absorption systems
at $z \sim 3$ if $z_{in} \la 5$, considering both the observational
and our theoretical uncertainties. 

\begin{figure}
\protect\centerline{
\epsfysize=2.5in\epsffile[20 155 570 700]
{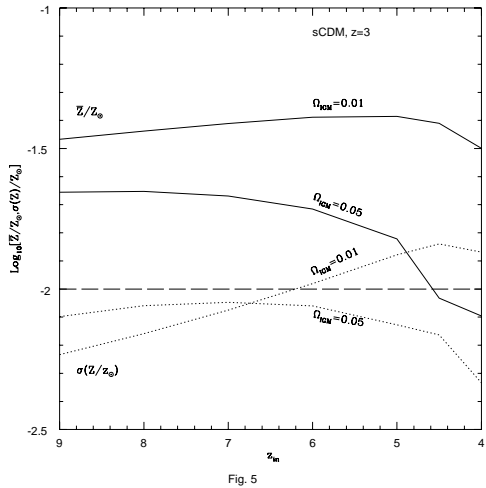}
}
\caption{The mean metallicity of the IGM (solid lines)
and the standard deviation (dotted lines) at $z=3$ are shown as
functions of $z_{in}$, for $\Omega_{IGM}=0.01$ and $0.05$. The
dashed line shows the reported metallicity of the $\lyal$ absorption
lines at $z=3$.}
\end{figure}

It should be noted here that the determination of metallicity in
the $\lyal$ absorption systems depends on the assumed value of the
ionization parameter $\Gamma$ --
the ratio of the density of ionizing photons to 
that of the particles in the $\lyal$ systems--which is uncertain. 
The resultant metallicity measured is then uncertain to about a factor
of two.
It is hoped that future observations (including those of
other elements than carbon) will give better values of the
metallicity and its scatter, so that the above model can be tested 
against the data.
In addition, abundance anomalies that are seen in the Ly$\alpha$ forest
may yield important information about the masses of the stars that
are producing the supernovae (Woosley \& Weaver 1992), which in turn 
will allow better nucleosynthesis yields to be used in predicting
the metallicities, in addition to information about $\Gamma$.  

\subsection{ Comparison with the metallicity of halo metal poor
stars}

It is well known that there exist in the Milky Way halo a
few stars with much lower Z
than we would expect given the metallicity of the initial gas cloud
out of which the Galaxy formed.
Six stars with $Z < 0.001$ Z$_{\odot}$ exist in the complete sample
of Ryan \& Norris 1991 (hereafter RN91 $-$
see their Fig.~5d); the lowest metallicity
star known is the carbon star G66$-$71, which has
$Z \sim 10^{-5.6}$Z$_{\odot}$ (Gass, Liebert \& Wehrse 1988).
The mean metallicity of halo stas is about
$Z \sim 10^{-1.8}$Z$_{\odot}$ (RN91).
Such large inhomogeneities might not have
existed
within the primordial gas cloud that was originally the
Milky Way  and are not required by this
model (however they are not inconsistent with observation;
RN91).
The most natural interpretation is that these stars are pre-galactic
and dynamically are associated with the dark matter, and
not the rest of the stellar halo (however, binary mass transfer effects,
like differential absorption of elements onto dust, might be
important in explaining the low metallicities of extremely
low-Z carbon stars like G66$-$71).
We now investigate this possibility.

A very approximate quantitative analysis is suggested.
Although the probability distribution functions calculated above
are not Gaussian, let us discuss the case of a Gaussian distribution,
for the sake of illustration.
In our model, where the IGM metallicity has a mean and spatial
variance of 0.01 Z$_{\odot}$, we would expect 18\% of all
pregalactic stars to have $Z < 0.001$ Z$_{\odot}$.
Here ``pregalactic'' refers to
stars that formed out of material processed only through
the phase described by this model and also to any stars
that might represent the low-mass tail of the IMF describied
in Section 3;
stars that formed from gas that had been enriched
further in smaller galaxies that were Milky Way
progenitors are excluded.
As 6 stars in the RN91 sample satisfy this constraint, we
would then estimate that 32 stars in their sample are
pregalactic of which 16 would have $Z < 0.01$ Z$_{\odot}$.
This is 16\% of the total number of stars (104) in the RN91 sample
with $Z < 0.01$ Z$_{\odot}$, which in turn suggests
16\% of the stars in the halo with $Z < 0.01$ Z$_{\odot}$ are pregalactic.
If one takes all the stars in the RN91 sample (372), then one
finds $32/372=9\%$ of all halo stars to be pregalactic.
The rest of the stars formed during collapse of the halo, or
in MW progenitors.

So it is natural in our model that these stars should exist.
Our model also predicts the existence of a few very low $Z$
stars like
G66-71 (although in this case binary mass transfer is the
more plausible explanation, particularly in light of the
inverted C/O abundance; Gass, Liebert \& Wehrse 1988).
Surveys of the low $Z$ halo stars, including both
proper motion surveys like RN91 and objective prism surveys,
are rapidly getting larger, and a metallicity probability
distribution function for the Galactic halo should soon
be available with much better statistics at the low $Z$
end than that in RN91.  Then the calculation in the
previous paragraph may be done rigorously.  This will
provide one of the strongest constraints on the model
presented here, and more generally on models that seek
to explain the origin of the metals in the IGM.

The pregalactic stars, as defined above, should have $r^{-2}$
density profiles, like the dark matter, not $r^{-3}$ profiles,
like most of the stellar halo (Rees 1997).
Recent analysis of point-sources in the Hubble Deep Field
(Elson, Santiago \& Gilmore 1996) suggest that if the
Milky Way has an extended stellar
halo, this extended halo must be very small compared to the
rest of the halo.  This suggests that the fraction of
halo stars that are pregalactic is very low, and that our
estimate earlier in this secion is too high.
This prediction will not, however, be testable for some time
because surveys of low $Z$ stars would need to be carried
out over regions well outside the solar neighborhood,
where most stars are very faint.
The best chance of making this measurement in the future is if
we can identify a population of
horizontal branch stars that are likely to be
pregalactic based on metallicity and abundance ratio
considerations.

Another potentially
powerful probe of the properties of
pregalactic stars involves
heavy neutron-capture element
(e.g.~strontium = Sr)
abundances
(see Truran 1981, also Magain 1989; Ryan, Norris, \& Bessell
1991).
In the scenario described by these authors, the first
generation of stars (i.e.~that described in this paper)
only synthesizes iron peak nuclei from the unenriched gas.
The second generation performs secondary neutron
capture.  Only third or later generation stars would
then have the heavy neutron-capture elements.
This would in turn
mean that any primordial stars should be
deficient in these elements.  So would be any second generation
stars, which are also among the ones we call ``pregalactic'' here.
The most detailed survey of Sr abundances comes from
Ryan, Norris, \& Beers (1996).
Lower metallicity stars
do tend to have less strontium.
However, the evidence suggests that very few indeed, if
any, are completely Sr
deficient.
Again, this means that the pregalactic
fraction of halo stars that we estimated
above is probably too high.  It also means that
the IMF in the galaxies in Section 3 is probably biassed
towards high masses (if it was Salpeter, then at
least 20\% of halo stars would have been pregalactic
(specifically first-generation);
Miralda-Escud\'e \& Rees 1997).
An important caveat is that accretion of material
(in this case Sr) from the ISM onto the atmospheres of
the halo stars is small
(see Yoshii 1981), so that the Sr-deficient
population we observe is the true primordial one.

\subsection{Constraints from the UV background radiation}

Gunn-Peterson tests show that
the IGM was highly ionized by $z \sim 5$.
Several models for the reionization of the IGM have been discussed in
the literature, especially those investigating  
photoionization by various sources 
(quasars, star forming young galaxies, decaying neutrinos). Shapiro
\& Giroux (1987; see also Giroux \& Shapiro 1996) claimed that 
ionizing photons from the observed quasar population were not enough
for reionizing the IGM. Miralda-Escud\'e \& Ostriker (1990) showed that
including the UV photons from the star forming galaxies at high redshift
would make the spectrum of the UV background radiation softer. 

The UV background radiation at high redshift has been estimated
using the proximity effect (Bajtlik, Duncan \& Ostriker 1988).
It is known that the intensity of the UV
background at the Lyman limit is $\sim 10^{-21 \pm 0.5}$ erg cm$^{-2}$
s$^{-1}$ Hz$^{-1}$ sr$^{-1}$ at 
$z \sim 2 \hbox{--} 3$, although the value at
higher redshifts remains uncertain (see, e.g., Bechtold 1994;
Cristiani \etal 1996). Gunn-Peterson tests for both H~I and He~II
atoms at high redshift give clues to the spectrum of the UV
background radiation. The recent observations of He~II absorption
at $z \sim 3$ 
(see e.g.~Songalia, Hu \& Cowie 
1995) have been used to further constrain the spectrum.
Madau \& Meiksin (1994) showed that the observed constraints allowed
several possible sources, ranging from only observed quasars to
a combination of quasars and star forming galaxies.

Star formation responsible for the 
galactic winds at high redshift discussed
above would necessarily produce UV photons, and might reionize the 
IGM at high redshift. Silk, Wyse \& Shields (1987) suggested that
early galactic winds from dwarf galaxies would enrich the IGM and
could also provide a soft UV background radiation. Miralda-Escud\'e
\& Ostriker (1990) noted that the intensity of the radiation is
consistent with a cosmic metallicity of $\sim 0.01$ solar. Giroux
\& Shapiro (1996) considered the problem of metal production in
galaxies responsible for photoionization of the IGM in detail. They
showed that the metal enrichment inside the galaxies could reach
$20\% \hbox{--} 30\%$ of solar metallicity by $z \sim 3$ if the
metals are not ejected from the parent galaxies.

Madau \& Shull (1996) considered the metal production rate in
star forming galaxies at high redshift.
They did not however consider physical processes that might distribute
the metals outside the parent galaxies; their models differ 
fundamentally from
ours in that they assume that the metals observed in the Ly$\alpha$
clouds at $z=3$ were made inside those clouds. 
They considered continuous
metal production in galaxies from $z \la 4$ and determined the
resulting UV background radiation at $z \sim 3$, after letting the
UV photons traverse the cloudy IGM.  For continuous star formation
from $z=4$, and for reaching a metallicity of $0.01$ solar in
the parent galaxies by $z=3$,
the resulting UV background radiation has an intensity (at the Lyman
limit) of $0.13 \times 10^{-21}$ erg cm$^{-2}$
s$^{-1}$ Hz$^{-1}$ sr$^{-1}$. 
This assumes that a fraction $f_{esc}=0.25$
of the UV photons escapes the parent galaxies. This value of the
intensity is small compared to that estimated from proximity effect
(see above).
The important concept stressed by these authors is that
$f_{esc}$ is {\bf very} poorly known (see also Leitherer et 
al.~1995, and the discussion of that paper in Madau \& Shull 1996,
for observational constraints).  In addition, the IGM itself
might have considerable opacity at $z > 5$ (no Gunn-Peterson
constraints exist), and this might also reduce the contribution of
the UV photons produced to the cosmic background.
  
In our model, the galaxies emit their 
UV photons at higher redshifts
than considered by Madau \& Shull (1996).  Given their
calculations, 
the contribution of these photons to the
UV background at $z \sim 3$
must be even less than produced in  
their $3 < z < 4$ continuous star-formation
model.  Given this, and the uncertainties above, 
it seems reasonable to state that the UV background we produce
is likely to be significantly smaller than what is observed
through the proximity effect.

\subsection{Can we see this process in progress?}

The brightest objects at $z_{in}$ will be the supernovae that produce
the metals here.  They will have $K \sim 27$ at $z=5$
(duration $\sim 1$ year) in the observer frame,
which is too faint to see from the ground
(Miralda-Escud\'e \& Rees, 1997).  

An interesting idea suggested by 
Miralda-Escud\'e \& Rees (1997) is that
a few of these high-$z$ supernovae will happen to fall behind
rich clusters of galaxies and so have their total flux amplified
by gravitational lensing.  The cross-section that they estimate
for magnifying a source by more than $A=100$ is about 10$^{-4}$
arcminutes (this is the minimum magnification for such an object
to be detected at present).  This suggests that the probability
of seeing such an object in any given rich
cluster at a given time is 0.01\%.
If we surveyed 100 clusters to $K=22$ every year, we would then
have a 1\% chance of finding one of these SN.  
As sensitive $K$-band detectors become available on large
telescopes, lower $A$ events become detectable and this project
becomes feasible (optical detection is unlikely because
observer-frame optical wavelengths probe rest-frame UV
wavelengths, where supernovae are intrinsically faint). 

The discovery of such an object should be straightforward
to confirm spectroscopically, unless its $z$ is so high
($z>10$) the strongest emission lines get redshifted too
far into the infrared, where such faint-object spectroscopy is
impossible. 
 
\subsection{What is the fate of the small galaxies?}

In our model,
the galaxies responsible for producing the metals
had their baryons blown out by the galactic winds
(Dekel \& Silk 1986).  Therefore,
if they survive to the present day, they must be almost
dark, unless they have undergone recent infall of gas from
the IGM and made more stars (this is the scenario
described by Silk, Wyse, \& Shields 1987).

A few almost dark, low-mass
galaxies are known in the Local
Group.  The most extreme examples are Carina
(Mateo et al.~1993), Draco, and Ursa Minor
(Pryor \& Kormendy 1990).
Galaxies like Draco and Ursa Minor
have stellar populations that are not self-gravitating
and are $>99$\% dark matter by mass within their core-fitting
radius.  Their total dark-matter fractions depend on the
mass distribution outside this radius, and will probably
be higher still.
These galaxies are the natural remnants of the process
described in this paper.  The few stars that they do have
probably result from later gas infall and
subsequent star-formation.
They do not have zero metallicity (Hodge 1989)
so do not represent the
low-mass tail of the IMF described earlier in this paper.
Very few low-mass first-generation stars, if any, exist in these
galaxies (a similar low-mass star deficient mass function has
been suggested for the Galactic halo; Nakamura
Kan-ya \& Nishi 1997).

But these galaxies do not exist in large number.
The galaxy luminosity function in the Local Group
(van den Bergh 1992) is certainly
not increasing rapidly at $M_B \sim -7$, the luminosity of
Draco \& Ursa Minor.

So where have the galaxies that produce the metals gone?
Many might have undergone considerable subsequent gas infall
(as argued in Silk, Wyse \& Shields 1987) and so be luminous
galaxies today.  They may even have merged with larger
galaxies, or even have been tidally disrupted.
Draco, Ursa Minor, and Carina, if they really are
remnants of the present model, would then be anomalous in
that they had not suffered any of these fates

An alternative is that more recent gas infall has not
happened in most remnants of the present process, and that
there exist many undetected dark galaxies today.
Draco and Ursa Minor would then be anomalously {\bf bright}
examples of these objects, where the little amount of
gas they picked up allowed them to form a few stars, which
in turn allowed them to be detected.

\subsection{The origin of the X-ray emitting gas in clusters}

Presumably the X-ray emitting gas that is now seen in clusters
has experienced the process described in this paper (although the
galaxies here are not the source of most of the ICM metals; see
Section 2).  In two previous papers (Trentham 1994, NC95), we
argued that dwarf galaxies might contribute much of the ICM.
In the context of the present model, much of the IGM might
well have been
processed in small galaxies, but these are very different systems
from the present-epoch cluster
dwarf galaxy population discussed in
those papers.  

\subsection{Concluding remarks}

To summarize, we have found that galactic winds
at high $z$ could have enriched the IGM to a mean metallicity of
$Z \sim 0.01 $Z$_{\odot}$
at $z \sim 3$,
with a standard deviation of the same
order, if $z_{in} \la 5$, and that this satisfies all the
observational constraints.

While our results show that these kinds of
processes can easily account for the observed IGM metallicity, we
needed to make a number of simplifying assumptions and adopt convenient
but plausible values for our model parameters. 
This work has been presented mostly in the spirit of a
plausibility argument, and we have not attempted a detailed
study of parameter space.
Numerical investigations will hopefully provide more insight
into this process and will allow an assessment of the 
permissible ranges for the input parameters in models like
the present one.  These will 
provide important constraints on galaxy formation theories.

Additional observations will also provide stronger constraints.
Specifically, measurements which should be particularly valuable
are (i) measurement of abundances ratios in the Ly$\alpha$
forest, (ii) the metallicity distribution function of the 
Galactic halo for a large sample of stars, and (iii) 
heavy neutron-capture elemental abundances for an equally large sample.

\bigskip

B.N. thanks Drs. M. Chiba, P. Biermann  and 
N.T. thanks Drs. L. Cowie, J. Norris, and M. Rees for stimulating
discussions. The authors also thank the anonymous referee for his comments.


\begin{thebibliography}{}

\bibitem[]{} Arnaud, M., Rothenflug, R., Boulade, O., Vigroux, L.,
\& Vangioni-Flam, E. 1992, A\&A, 254, 49
\bibitem[]{} Babul, A., \& Rees, M. J. 1992, MNRAS, 255, 346
\bibitem[]{} Bajtlik, S., Duncan, R.~C., \& Ostriker, J.~P. 1988, ApJ, 327,
520
\bibitem[]{} Bechtold, J. 1994, ApJS, 91, 1
\bibitem[]{} Blumenthal, G. R.,
Faber, S. M., Primack, J. R., \& Rees, M. J., 1984, Nature, 311, 517 
\bibitem[]{} Broadhurst, T. J, Ellis, R. S, \& Shanks, T. 1988, 
MNRAS, 235, 827
\bibitem[]{} Bunn, E. F., \& White, M. 1996, preprint (astro-ph.9607060v2)
\bibitem[]{} Carr, B.~J., Bond, J.~R., \& Arnett, W.~A. 1984, ApJ, 277, 445
\bibitem[]{} Couchman, H.~M.~P., \& Rees, M.~J. 1986, MNRAS, 221, 53
\bibitem[]{} Cristiani, S., D'Odorico, S., Fontana, A., Giallongo, E., \&
Savaglio, S. 1995, MNRAS, 273, 1016
\bibitem[]{} Cowie, L. L., Hu, E. M., \& Songaila, A. 1995, Nature, 377, 603
\bibitem[]{} Cowie, L. L., Songaila, A., \& Hu, E. M. 1991, Nature, 354, 460
\bibitem[]{} Cowie, L. L., Songaila, A., Kim, T. S., \& Hu, E. M. 1995, 
AJ, 109, 1522
\bibitem[]{} Dekel, A. \& Silk, J. 1986,  ApJ, 303, 39
\bibitem[]{} Efstathiou, G. 1992, MNRAS, 256, 43p 
\bibitem[]{} Eggen, O. J., Lynden-Bell, D., \& Sandage, A. R. 1962,
ApJ, 136, 748 
\bibitem[]{} Elson, R. A. W., Santiago, B. X. \& Gilmore, G. F.
1996, New Astron., 1, 1
\bibitem[]{} Fukugita, M., \& Kawasaki, M. 1994, MNRAS, 269, 563
\bibitem[]{} Gass, H., Liebert, J., \& Wehrse, R. 1988, A\&A, 189, 194
\bibitem[]{} Giroux, M. L., \& Shapiro, P. R. 1996, ApJS, 102, 191
\bibitem[]{} Gnedin, N. Y., \& Ostriker, J. P. 1992, ApJ, 400, 1
\bibitem[]{} Gnedin, N. Y., \& Ostriker, J. P. 1997, preprint 
(astro-ph/9612127) 
\bibitem[]{} Haiman, Z., Rees, M. J., \& Loeb, A. 1996, 
preprint (astro-ph/9608130) 
\bibitem[]{} Heckman, T. M., Armus, L., \& Miley, G. K. 1990, ApJS, 
74,833
\bibitem[]{} Hodge, P.~W.~1989, ARAA, 27, 139 
\bibitem[]{} Leitherer, C., Ferguson, H. C., Heckman, T. M.,
Lowenthal, J. D. 1995, ApJ, 454, L19
\bibitem[]{} Loewenstein, M., \& Mushotzky, R. F. 1996, ApJ, 466, 695
\bibitem[]{} Lu, L., Sargent, W.L.W., \& Barlow, T.A. 1996, preprint
  (astro-ph/9606044)
\bibitem[]{} Madau, P., \& Meiksin, A. 1994, ApJ, 433, L53
\bibitem[]{} Madau, P., \& Shull, J. M. 1996, ApJ, 457, 551
\bibitem[]{} Magain, P. 1989, A\&A, 209, 211
\bibitem[]{} Mateo, M., Olszewski, E. W., Pryor, C., Welch, D. L., \&
Fischer, P. 1993, AJ, 105, 510
\bibitem[]{} Matteucci, F., \& Gibson, B. K. 1995, A\&A, 304, 11 
\bibitem[]{} Meyer, D. M., \& York, D. G. 1987, ApJ, 315, L5
\bibitem[]{} Miralda-Escud\'e, J., \& Ostriker, J. P. 1990, ApJ, 350, 1
\bibitem[]{} Miralda-Escud\'e, J., \& Rees, M. J. 1997, ApJ, in press
(astro-ph/9701093)
\bibitem[]{} Mushotzky, R. F., Loewenstein, M., Arnaud, K. A.,
Tamura, T., Fukazawa, Y., Matsushita, K., Kikuchi, K., \&
Hatsukade, I. 1996, ApJ, 466, 686 
\bibitem[]{} Nakamura, T., Kan-Ya, Y. \& Nishi, R. 1997, ApJ, 473, L99
\bibitem[]{} Nath, B. B., \& Chiba, M. 1995, ApJ, 454, 604 (NC95)
\bibitem[]{} Peacock, J. S., \& Heavens, A. F. 1990, MNRAS, 243, 133
\bibitem[]{} Press, W.H., \& Schechter, P. 1974, ApJ, 187, 425
\bibitem[]{} Pryor, C., \& Kormendy, J., 1990, AJ, 100, 127
\bibitem[]{} Rauch, M., Haehnelt, M. G. \& Steinmetz, M. 1996, preprint
(astro-ph/9609083)
\bibitem[]{} Rees, M. J. 1997, in $HST$ and the Distant Universe,
ed. A.~Aragon-Salamanca, in press 
\bibitem[]{} Ryan, S. G., \& Norris, J. E. 1991, AJ, 101, 1865 (RN91) 
\bibitem[]{} Ryan, S. G., Norris, J. E., \& Beers, T. C. 1996, ApJ, 
in press
\bibitem[]{} Ryan, S. G., Norris, J. E., \& Bessell, M. S. 1991,
AJ, 102, 303
\bibitem[]{} Sanders, D. B., \& Mirabel, I. F. 1996, ARAA, 34, 749 
\bibitem[]{} Shapiro, P. R., \& Giroux, M. L. 1987, ApJ, 321, L107
\bibitem[]{} Silk, J., Wyse, R. F. G., \& Shields, G. A. 1987,
 ApJ, 322, L59
\bibitem[]{} Songaila, A., \& Cowie, L. L. 1996, AJ, 112, 335
\bibitem[]{} Songaila, A., Hu, E. M.,  \& Cowie, L. L. 1995, 
Nature, 375, 124
\bibitem[]{} Tegmark, M., Silk, J., \& Evrard, A. 1993, ApJ, 417, 54
\bibitem[]{} Tegmark, M., Silk, J., Rees, M. J., Blanchard, A.,
Abel, T., \& Palla, F. 1997, ApJ, 474, 1
\bibitem[]{} Thoul, A. A., \& Weinberg, D. H. 1995, ApJ, 442, 480  
\bibitem[]{} Trentham, N. 1994, Nature, 372, 157 
\bibitem[]{} Truran, J. W. 1981, A\&A, 97, 391 
\bibitem[]{} Tytler, D. 1987, ApJ, 321, 49
\bibitem[]{} Tytler, D., Fan, X.-M., Burles, S., Cottrell, L., Davis, C.,
 Kirkman, D., \& Zuo, L. 1995,, in QSO Absorption Lines, ed. G. Meylan
 (Berlin: Springer-Verlag), p. 289
\bibitem[]{} van den Bergh, S.~1992, A\&A, 264, 75 
\bibitem[]{} Voinovich, P. A. \& Chernin, A. D. 1995, Astronomy Lett.,
21, 835 (translated from Pis'ma Astron. Zh. 1995, 21, 926)
\bibitem[]{} Voit, G. M. 1996, ApJ, 465, 548 
\bibitem[]{} White, S. D. M., \& Frenk, C. 1991, ApJ, 379, 52 
\bibitem[]{} Womble, D. S, Sargent, W. L. W., \& Lyons, R. S. 1996, in
  Cold gas at High Redshift, eds. M. Bremer, H. Rottgering,
  P. van der Werf, C. Carilli (Kluwer), in press
\bibitem[]{} Woosley, S. E., \& Weaver, T. A. 1982, in  
Supernovae: A survey of current research, eds. M. J. Rees,
R. J. Stoneham (Dordrecht: Reidel), p.~79
\bibitem[]{} Yoshii, Y. 1981, A\&A, 97, 280 
\bibitem[]{} Yoshii, Y., \& Arimoto, N. 1987, A\&A, 188, 13
\bibitem[]{} Yoshioka, S. \& Ikeuchi, S. 1990, ApJ, 360, 352


\end{thebibliography}
\end{document}